\documentclass[twocolumn,graphicx,pra]{revtex4}
\usepackage{amssymb}
\usepackage{epsfig}
\usepackage{bm}
\usepackage{amsmath}
\usepackage{graphicx}

\begin{document}

\title{Excitation, two-center interference and the orbital geometry in
laser-induced nonsequential double ionization of diatomic molecules }
\author{T. Shaaran, B.B. Augstein and C. Figueira de Morisson Faria}
\affiliation{Department of Physics and Astronomy, University College London, Gower
Street, London WC1E 6BT, United Kingdom}
\date{\today}

\begin{abstract}
We address the influence of the molecular orbital geometry and of
the molecular alignment with respect to the laser-field polarization
on laser-induced nonsequential double ionization of diatomic
molecules for different molecular species, namely $\mathrm{N}_2$ and
$\mathrm{Li}_2$. We focus on the recollision excitation with
subsequent tunneling ionization (RESI) mechanism, in which the first
electron, upon return, promotes the second electron to an excited
state, from where it subsequently tunnels. We show that the
electron-momentum distributions exhibit interference maxima and
minima due to the electron emission at spatially separated centers.
We provide generalized analytical expressions for such maxima or
minima, which take into account $s$ $p$ mixing and the orbital
geometry. The patterns caused by the two-center interference are
sharpest for vanishing alignment angle and get washed out as this
parameter increases. Apart from that, there exist features due to
the geometry of the lowest occupied molecular orbital (LUMO), which
may be observed for a wide range of alignment angles. Such features
manifest themselves as the suppression of probability density in
specific momentum regions due to the shape of the LUMO wavefunction,
or as an overall decrease in the RESI yield due to the presence of
nodal planes.

\end{abstract}

\maketitle

\subsection{Introduction}

Strong-field phenomena such as high harmonic generation (HHG) or
above-threshold ionization (ATI) have been used as tools for the attosecond
imaging of molecular orbitals \cite{imaging}, for probing the structural
changes in molecules with attosecond precision and for studying quantum
interference effects due do photoelectron or high-harmonic emission at
spatially separated centers \cite{Probing}. This has been made possible due
to the fact that both phenomena are caused by the rescattering or
recombination of an electron with its parent molecule, which, for typical
intense lasers, occur within hundreds of attoseconds. The simplest targets
for which this interference can be studied are diatomic molecules, which can
be viewed as the microscopic counterpart of a double-slit experiment \cite%
{doubleslit}.

Potentially, laser-induced nonsequential double ionization (NSDI)
can also be employed for probing molecular orbitals since
laser-induced recollision plays an important role in this case. In
NDSI the returning electron rescatters inelastically with its parent
ion, or molecule, giving part of its kinetic energy to a second
electron. This electron can be released in
the continuum either through electron-impact ionization \cite%
{A.Becker,Carla1,Carla2,M.Lein,X.Liu,prauzner,A.Staudte,Emanouil,Bondar} or
recollision excitation with subsequent tunneling ionization (RESI) \cite%
{RESI1,RESI2}. The former recollision mechanism happens when the
first electron, upon return, gives enough energy to the second
electron of the target so that it can overcome the second ionization
potential and reach the continuum. The latter recollision mechanism
happens when the first electron, upon return, gives just enough
energy to the second electron so that it can be promoted to an
excited bound state, from where it subsequently tunnels.

In principle, NSDI exhibits several advantages, with regard to ATI
or HHG. First, it allows one to extract more dynamic information
about the system, as the type of electron-electron interaction can
be identified in the electron-momentum distributions
\cite{Carla1,Carla2,Emanouil}. Furthermore, different rescattering
mechanisms, such as electron-impact ionization or RESI, populate
different regions in momentum space and hence can also be traced
back from such distributions \cite{routes}. Apart from that, events
happening at different half cycles of the driving field can be
mapped into different momentum regions. Concrete examples are NSDI
with few-cycle pulses \cite{fewcycle}, which lead to asymmetric
electron-momentum
distributions, and individual processes in NSDI of diatomic molecules \cite%
{F2009}. Finally, electron-electron correlation is at the essence of this
phenomenon and cannot be ignored. In contrast, for high-order harmonic
generation one may, to first approximation, only consider a single active
electron and the highest occupied molecular orbital. In fact, only very
recently have multiple orbitals and electron-electron correlation been
incorporated in the modeling of molecular high-order harmonic generation
\cite{CarlaBrad,Patchkovskii,Lin,Olga2009,Haessler2010}.

 For the
above-mentioned reasons, NSDI of molecules is being increasingly
investigated since the past few years. In fact, there has been
experimental evidence that the orbital symmetry \cite{NSDIsymm} and
the alignment angle \cite{NSDIalign} affect the shapes of the
electron momentum distributions. Since then, many theoretical
studies have also been performed for molecules, involving, for
instance, classical trajectory methods \cite{classical}, the
numerical solution of the time-dependent Schr\"{o}dinger equation in
reduced-dimensionality models \cite{TDSEmol}, and semi-analytical
approaches based on the strong-field approximation
\cite{Smatrix_mol,NSDIInterference,F2009}. Semi-analytical models for NSDI in molecules,
however, focus on the electron-impact ionization rescattering mechanism. 

For instance, in our previous paper \cite{NSDIInterference} we
addressed the influence of the orbital symmetry and the molecular
alignment with respect to the laser-field polarization on NSDI of
diatomic molecules for the electron-impact ionization mechanism. We
showed that the electron momentum distribution exhibit interference
maxima and minima due to electron
emission at spatially separated centers. Such fringes were positioned at $%
p_{1||}+p_{2||}=const.$, i.e., parallel to the anti diagonal of the
plane spanned by the electron momentum components $p_{n\parallel }$
$n=1,2$ parallel to the laser-field polarization. They were sharpest
if the molecule was aligned along the direction of the field, i.e.,
for vanishing alignment angle. As this angle increased, the fringes
got increasingly blurred until they were completely washed out for
perpendicular alignment.

Apart from that, recently, several studies have found that the core
dynamics, in particular excitation, is important for high-harmonic
generation in molecules \cite{Olga2009,Haessler2010}, and in particular for attosecond imaging of matter. We expect this also
to be the case for nonsequential double ionization. For that reason, in the past few years, we have focused on the RESI mechanism. We have shown that the shape of the electron momentum distributions depends very
strongly on the initial and excited bound states of the second electron \cite{Shaaran,RESIM}, in fact far more critically than for electron-impact
ionization \cite{Carla2}. If this is the case already for single atoms, one
expects this dependence to be even more critical for molecules.

For RESI, we expect the electron momentum distributions to be
affected very strongly by the geometry of the bound-state
wavefunctions, not only because the excitation process strongly
depends on them, but also due to the fact that the second electron
is reaching the continuum by tunneling. It is by now well known that
this ionization mechanism is strongly influenced by the presence of
nodal planes or the directionality of a particular molecular
orbital. For instance, for HHG the nodal plane of a $\pi $ state
suppresses tunnel ionization when it
coincides with the polarization axis
(see, e.g., \cite{Olga2009,CarlaBrad,Haessler2010,Dejan2009,eliot,BradCarla}). 

In the present paper, we perform a systematic analysis of
quantum-interference effect in NSDI of diatomic molecules
considering the RESI mechanism. We construct a semi-analytical
model, based on the strong-field approximation (SFA), in which an
electron tunnels from the HOMO of a neutral molecule and rescatters
with the HOMO of its singly ionized counterpart. Thereby, we assume
that the second electron is excited to the lowest unoccupied
molecular orbital (LUMO). We investigate the influence of such
orbitals and of the alignment angle on the NSDI electron momentum
distributions. Specifically we choose species for which these
orbitals have different geometries and parities. Furthermore, we
address the question of whether well-defined interference patterns
such as those observed in ATI or HHG computations may also be
obtained for NSDI in the context of the RESI mechanism, and, if so,
under which conditions. These are complementary studies to those
performed in our recent work on RESI \cite{Shaaran,RESIM}, where we
show that, for single atoms, the shapes of the electron momentum
distributions carry information about the bound state from which the
second electron leaves and the state to which it is excited.

This paper is organized as follows. In Sec.~\ref{transampl} we
discuss the expression for the RESI transition amplitude, including
its general expression \ (Sec.~\ref{generalexpr}), the saddle-point
equations obtained from it (Sec.~\ref{saddle}) and the specific
prefactors for a diatomic molecule using Gaussian orbital basis sets
(Sec.~\ref{prefactors}). At the end of this section,
(Sec.~\ref{InterferenceCondition}) we derive a general\ two-center
interference condition for the RESI mechanism. Subsequently, in
Sec.~\ref{results}, we compute electron momentum distributions, with
emphasis on the two-center interference (Sec.~\ref{alignment}), and
the influence of different molecular orbitals (Sec.~\ref{orbits}).
Finally, in Sec.~\ref{conclusions}, we state the main conclusions to
be drawn from this work.

\section{Strong-field approximation transition amplitude}

\label{transampl}

\subsection{General expressions}

\label{generalexpr}\qquad

The SFA transition amplitude describing the RESI mechanism reads (for
details on the derivation see \cite{Shaaran}).

\begin{eqnarray}
M(\mathbf{p}_{n},t,t^{\prime },t^{^{\prime \prime }}) &=&\int_{-\infty
}^{\infty }\hspace*{-0.2cm}dt\hspace*{-0.1cm}\int_{-\infty }^{t}\hspace*{%
-0.3cm}dt^{\prime }\hspace*{-0.1cm}\int_{-\infty }^{t^{\prime }}\hspace*{%
-0.3cm}dt^{^{\prime \prime }}\hspace*{-0.1cm}\int d^{3}k  \notag \\
&&V_{\mathbf{p}_{2}e}V_{\mathbf{p}_{1}e,\mathbf{k}g}V_{\mathbf{k}g}e^{iS(%
\mathbf{p}_{n},\mathbf{k},t,t^{\prime },t^{\prime \prime })},  \label{Mp}
\end{eqnarray}

with the action

\begin{eqnarray}
S(\mathbf{p}_{n},\mathbf{k},t,t^{\prime },t^{^{\prime \prime }})
&=&-\int_{t}^{\infty }\hspace{-0.1cm}\frac{[\mathbf{p}_{2}+\mathbf{A}(\tau
)]^{2}}{2}d\tau  \notag \\
&&-\int_{t^{^{\prime }}}^{\infty }\hspace{-0.1cm}\frac{[\mathbf{p}_{1}+%
\mathbf{A}(\tau )]^{2}}{2}d\tau  \notag \\
&&-\int_{t^{^{\prime }}}^{t^{^{\prime \prime }}}\hspace{-0.1cm}\frac{[%
\mathbf{k}+\mathbf{A}(\tau )]^{2}}{2}d\tau  \notag \\
&&+E_{1g}t^{^{\prime \prime }}+E_{2g}t^{^{\prime }}+E_{2e}(t-t^{^{\prime }})
\label{singlecS}
\end{eqnarray}

and the prefactors%
\begin{eqnarray}
V_{\mathbf{k}g} &=&\left\langle \mathbf{\tilde{k}}(t^{\prime \prime
})\right\vert V\left\vert \psi _{g}^{(1)}\right\rangle =\frac{1}{(2\pi
)^{3/2}}  \notag \\
&&\times \int d^{3}r_{1}V_{0}(\mathbf{r}_{1})\exp [-i\mathbf{\tilde{k}}%
(t^{\prime \prime })\cdot \mathbf{r}_{1}]\psi _{g}^{(1)}(\mathbf{r}_{1})
\label{Vkg}
\end{eqnarray}

\begin{eqnarray}
V_{\mathbf{p}_{1}e\mathbf{,k}g} &=&\left\langle \mathbf{\tilde{p}}_{1}\left(
t^{\prime }\right) ,\psi _{e}^{(2)}\right\vert V_{12}\left\vert \mathbf{%
\tilde{k}}(t^{\prime }),\psi _{g}^{(2)}\right\rangle =\frac{1}{(2\pi )^{3}}
\notag \\
&&\times \int \int d^{3}r_{2}d^{3}r_{1}\exp [-i(\mathbf{p}_{1}-\mathbf{k}%
)\cdot \mathbf{r}_{1}]  \notag \\
&&\times V_{12}(\mathbf{r}_{1,}\mathbf{r}_{2})[\psi _{e}^{(2)}(\mathbf{r}%
_{2})]^{\ast }\psi _{g}^{(2)}(\mathbf{r}_{2})  \label{Vp1e,kg}
\end{eqnarray}

and \ \
\begin{eqnarray}
V_{\mathbf{p}_{2}e} &=&\left\langle \mathbf{\tilde{p}}_{2}\left( t\right)
\right\vert V_{\mathrm{ion}}\left\vert \psi _{e}^{(2)}\right\rangle =\frac{1%
}{(2\pi )^{3/2}}  \notag \\
&&\times \int d^{3}r_{2}V_{\mathrm{ion}}(\mathbf{r}_{2})\exp [-i\mathbf{%
\tilde{p}}_{2}(t)\cdot \mathbf{r}_{2}]\psi _{g}^{(2)}(\mathbf{r}_{2}).
\label{Vp2e}
\end{eqnarray}

Eq. (\ref{Mp}) describes the physical process in which, at a time
$t^{\prime \prime },$\ the first electron tunnels from a bound state
$|\psi _{g}^{(1)}>$ \ into a Volkov state
$|\mathbf{\tilde{k}}(t^{\prime })>$. Then the released electron
propagates in the continuum from $t^{\prime \prime }$ to $t^{\prime
}$ , and it is driven towards its parent molecule. Upon return, the
electron scatters inelastically with the core at $t^{\prime }$ and,
through the interaction $V_{12},$ promotes the second electron from
the bound state $|\psi _{g}^{(2)}>$  to the excited state $|\psi
_{e}^{(2)}>$. Finally, at a
later time $t,$ the second electron, initially in a bound excited state $%
|\psi _{e}^{(2)}>,$ is released by tunneling ionization into a Volkov state $%
|\mathbf{\tilde{p}}_{2}\left( t\right) >$. In the above-stated equations, $%
E_{ng}$ $(n=1,2)$ are the ionization potentials of the ground state, $E_{ne}$
$(n=1,2)$ denote the absolute values of the excited-state energies and the
potentials $V_{0}(\mathbf{r}_{1})$ \ and $V_{\mathrm{ion}}(\mathbf{r}_{2})$
\ correspond to the neutral molecule and the singly ionized molecular
species, respectively. Here, the final electron momenta are described by $%
\mathbf{p}_{n}(n=1,2).$ All the information about the binding potentials
viewed by the first and second electrons and the electron-electron
interaction are embedded in the form factors (\ref{Vkg}) , (\ref{Vp2e}) and (%
\ref{Vp1e,kg}) respectively. Assuming that the electron-electron interaction
depends only on the difference between the two electron coordinates, i.e.,
if $V_{12}(\mathbf{r}_{1,}\mathbf{r}_{2})=V_{12}(\mathbf{r}_{1-}\mathbf{r}%
_{2}),$ Eq. (\ref{Vp1e,kg}) may be rewritten as
\begin{eqnarray}
V_{\mathbf{p}_{1}e\mathbf{,k}g} &=&\frac{V_{12}(\mathbf{p}_{1}-\mathbf{k})}{%
(2\pi )^{3/2}}  \notag \\
&&\times \int d^{3}r_{2}e^{-i(\mathbf{p}_{1}-\mathbf{k})\cdot \mathbf{r}%
_{2}}[\psi _{e}^{(2)}(\mathbf{r}_{2})]^{\ast }\psi _{g}^{(2)}(\mathbf{r}%
_{2}),  \label{resc1st}
\end{eqnarray}%
with
\begin{equation}
V_{12}(\mathbf{p}_{1}-\mathbf{k})=\frac{1}{(2\pi )^{3/2}}\int d^{3}re^{-i(%
\mathbf{p}_{1}-\mathbf{k})\cdot \mathbf{r}}V_{12}(\mathbf{r})
\end{equation}%
and $\mathbf{r=r}_{1-}\mathbf{r}_{2}.$

Within the framework of the SFA these prefactors are gauge
dependent. Specifically, in the length gauge
$\mathbf{\tilde{p}}_{n}\left( \tau \right)
=\mathbf{p}_{n}+\mathbf{A}(\tau )$ and $\mathbf{\tilde{k}}(\tau )=\mathbf{k}+%
\mathbf{A}(\tau )(\tau =t^{\prime },t^{\prime \prime }),$ while in
the velocity gauge $\mathbf{\tilde{p}}_{n}\left( \tau \right)
=\mathbf{p}_{n}$ and $\mathbf{\tilde{k}}(\tau )=\mathbf{k}.$ This is
due to the fact that the gauge transformation cancels out with the
minimal coupling in the latter case. In practice, however, for the
specific situation addressed in this work, both gauges lead to very
similar results. This happens as the above-stated phase differences
will cancel out in $V_{\mathbf{p}_{1}e,\mathbf{k}g}$, and, in
$V_{\mathbf{p}_{2}e}$, $\mathbf{A}(t)\simeq 0$ for the parameter
range of interest (for more details see \cite{RESIM}). In the
following, unless strictly necessary, we will drop the time
dependence in $\mathbf{\tilde{p}}_{n}\left( \tau \right) .$

\subsection{Saddle-point analysis}

\label{saddle}

Subsequently, the transition amplitude (\ref{Mp}) is solved employing
saddle-point methods. For that purpose, one must find the coordinates $%
(t_{s},t_{s}^{\prime },t_{s}^{\prime \prime },\mathbf{k}_{s})$ for which $S(%
\mathbf{p}_{n},\mathbf{k},t,t^{\prime },t^{\prime \prime })$ is stationary,
i.e., for which the conditions $\partial _{t}S(\mathbf{p}_{n},\mathbf{k}%
,t,t^{\prime },t^{\prime \prime })=\partial _{t^{\prime }}S(\mathbf{p}_{n},%
\mathbf{k},t,t^{\prime },t^{\prime \prime })=\partial _{t^{^{\prime \prime
}}}S(\mathbf{p}_{n},\mathbf{k},t,t^{\prime },t^{\prime \prime })=\mathbf{0}$ and $%
\partial _{\mathbf{k}}S(\mathbf{p}_{n},\mathbf{k},t,t^{\prime },t^{\prime
\prime })=0$ are satisfied. This leads to the equations

\begin{equation}
\left[ \mathbf{k}+\mathbf{A}(t^{\prime \prime })\right] ^{2}=-2E_{1g},
\label{saddle1}
\end{equation}%
\begin{equation}
\mathbf{k=}-\frac{1}{t^{\prime }-t^{\prime \prime }}\int_{t^{\prime \prime
}}^{t^{\prime }}d\tau \mathbf{A}(\tau )  \label{saddle2}
\end{equation}%
\begin{equation}
\lbrack \mathbf{p}_{1}+\mathbf{A}(t^{\prime })]^{2}=\left[ \mathbf{k}+%
\mathbf{A}(t^{\prime })\right] ^{2}-2(E_{2g}-E_{2e}).  \label{saddle3}
\end{equation}

\bigskip and%
\begin{equation}
\lbrack \mathbf{p}_{2}+\mathbf{A}(t)]^{2}=\mathbf{-}2E_{2e},  \label{saddle4}
\end{equation}%
which, as discussed below, provide additional physical insight into the
problem.

The saddle-point Eq. (\ref{saddle1}) gives the conservation of
energy when
the first electron tunnel ionized at a time $\ t^{\prime \prime }$. Eq. (\ref%
{saddle2}) constraints the intermediate momentum $\mathbf{k}$ of the first
electron and it makes sure the electron returns to the side of its release,
which lies at the geometrical center of molecule. Eq. (\ref{saddle3})
expresses the conservation of energy at\ a time $t^{\prime }$, when the
first electron rescatters inelastically with its parent ion, and gives part
of its kinetic energy $E_{\mathrm{ret}}(t^{\prime })=\left[ \mathbf{k}+\mathbf{A}%
(t^{\prime })\right] ^{2}/2$ to the core to excite the second electron from
a state with energy $E_{2g}$ to a state with energy $E_{2e}$. Immediately
after rescatering the first electron reaches the detector with momentum $%
\mathbf{p}_{1}$. Finally, Eq. (\ref{saddle4}) describes the fact that the
second electron tunnels at time $t$ from an excited excited state $E_{2e}$
and reaches the detector with momentum $\mathbf{p}_{2}.$ As a consequence of
the fact that tunneling has no classical counterpart, these equations
possess no real solutions (for more details see \cite{Shaaran}).

\subsection{Molecular prefactors}

\label{prefactors}

In this work, we consider that all molecular orbitals are frozen
apart from the HOMO and the LUMO. We also assume frozen nuclei and a
linear combination of atomic orbitals (LCAO) to construct
approximate wave functions for the active orbitals. This implies
that the molecular
bound-state wave function for each electron reads%
\begin{equation}
\psi ^{(n)}(\mathbf{\mathbf{r}}_{n})=\sum_{\alpha }c_{\alpha }[\phi _{\alpha
}^{(n)}(\mathbf{r}_{n}+\mathbf{R}/2)+(-1)^{l_{\alpha }+\lambda _{\alpha
}}\phi _{\alpha }^{(n)}(\mathbf{r}_{n}-\mathbf{R}/2)]  \label{Wave Function}
\end{equation}%
where $R$ and $l_{\alpha }$ denote the internuclear separation and the
orbital quantum numbers, respectively. The index $n=1,2$ refers to the
electron in question.$\ $\ The index $\lambda _{\alpha }=0$ applies to
gerade symmetry and $\lambda _{\alpha }=1$ to ungerade symmetry. The binding
potential of this molecule, as seen by each electron, is given by
\begin{equation}
V_{\varkappa }(\mathbf{r}_{n})=\mathcal{V}_{\varkappa }(\mathbf{r}_{n}-%
\mathbf{R}/2)+\mathcal{V}_{\varkappa }(\mathbf{r}_{n}+\mathbf{R}/2)
\label{binding potential}
\end{equation}%
where the subscript $\varkappa =0$ or ion refers either to the neutral
molecule or to its ionic counterpart, respectively, and $\mathcal{V}%
_{\varkappa }(\mathbf{r}_{n})=Z_{\mathrm{eff}}/r_{n}$ is the
potential at each center in the molecule. Thereby,
$Z_{\mathrm{eff}}$ is the effective core charge as seen by each of
the two active electrons.

In this paper the wave function $\phi _{\alpha }^{(n)}$ is approximated by a
Gaussian basis set,
\begin{equation}
\phi _{\alpha
}^{(n)}(\mathbf{r}_{n})=\sum_{j}b_{j}^{(n)}x^{l_{\alpha
}}y^{l_{\alpha }}z^{l_{\alpha }}\exp [-\zeta _{j}\mathbf{r}^{2}]
\label{Gaussian basis}
\end{equation}

The coefficients $b_{j}$ and $c_{\alpha }$ and the exponents $\zeta _{j}$
can be extracted either from existing literature or from quantum chemistry
codes. We compute these coefficients using GAMESS-UK \cite{GAMESS}. In our
basis set, we took only $s$ and $p$ states. This means that, in all the
expressions that follow, $l_{\alpha }$ and $l_{\beta }$ are either $0$ or $%
1. $

The above-stated assumptions lead to the form factors

\begin{eqnarray}
V_{\mathbf{p}_{1}e\mathbf{,k}g} &=&\frac{V_{12}(\mathbf{p}_{1}-\mathbf{k})}{%
(2\pi )^{3/2}}\sum_{\alpha }\sum_{\beta }[e^{i(\mathbf{p}_{1}-\mathbf{k}%
)\cdot \mathbf{\mathbf{R}}/2}  \notag \\
&&+(-1)^{l_{\alpha }+l_{\beta }+\lambda _{\alpha }+\lambda _{\beta }}e^{-i(%
\mathbf{p}_{1}-\mathbf{k})\cdot \mathbf{\mathbf{R}}/2}]\mathcal{I}_{1},
\label{vp1ke1}
\end{eqnarray}

where

\begin{equation}
\mathcal{I}_{1}=\int d^{3}r_{2}e^{-i(\mathbf{p}_{1}-\mathbf{k})\cdot \mathbf{%
r}_{2}}\phi _{\alpha }^{(2)}(\mathbf{r}_{2})^{^{\ast }}\phi _{\beta }^{(2)}(%
\mathbf{r}_{2})
\end{equation}

\bigskip and

\begin{equation}
V_{\mathbf{p}_{2}e}=\frac{4\pi }{(2\pi )^{3/2}}\sum_{\alpha }\left[ e^{i%
\mathbf{\tilde{p}}_{2}\cdot \mathbf{R}/2}+(-1)^{l_{\alpha }+\lambda _{\alpha
}}e^{-i\mathbf{\tilde{p}}_{2}\cdot \mathbf{R}/2}\right] \mathcal{I}_{2},
\label{vp2e1}
\end{equation}

\bigskip where $\ \ \ $%
\begin{equation}
\ \ \ \ \ \ \ \ \mathcal{I}_{2}=\int d^{3}r_{2}\mathcal{V}_{0}(\mathbf{r}%
_{2})e^{-i\mathbf{\tilde{p}}_{2}\cdot \mathbf{r}_{2}}\phi _{\alpha }^{(2)}(%
\mathbf{r}_{2}).
\end{equation}

In general, the form factor (\ref{Vkg}) does not affect the shape of the
electron-momentum distributions. This is particularly true when the first
electron tunnels from an orbital with no nodal planes, such as a $\sigma
_{g} $ orbital \cite{NSDIInterference}. However, one has to be careful when
the electron tunnels from any orbital with at least one nodal plane, such as
a $\pi $ orbital, as this would lead to a suppression of ionization for
specific alignment angles.

In the following, we will write the above-stated equations as functions of
the electron-momentum components $p_{n\parallel }$ and $\mathbf{p}_{n\perp }$
parallel and perpendicular to the laser-field polarization. Physically, we
are investigating a diatomic molecule whose main axis is rotated of an angle $%
\theta $ with respect to the direction of the laser-field
polarization. Hence, we are dealing with two frames of reference,
i.e., the molecular frame of reference and the laser field frame of
reference. The electron momenta in terms of their parallel and
perpendicular components with regard to the laser-field polarization
read
\begin{equation}
\mathbf{p}_{n}=p_{n||}\hat{e}_{z^{\prime }}+p_{n\perp }\cos \varphi \hat{e}%
_{x^{\prime }}+p_{n\perp }\sin \varphi \hat{e}_{y^{\prime }},
\end{equation}%
where we assumed that the laser field is polarized along the $z^{\prime }$
axis, the coordinates $x^{\prime }$ and $y^{\prime }$ define the plane
perpendicular to the laser-field polarization and $\varphi $ is the
azimuthal angle. In order, however, to compute the momentum-space
wavefunctions for this molecule, we need the momentum coordinates in the
frame of reference of the molecule. The molecular coordinates $x,y$ and $z$
can be obtained by a coordinate rotation around the $x$ axis. In this case,
the momenta of the electrons in terms of parallel and perpendicular
components in this latter frame of reference will be

\begin{figure}[tpb]
\begin{center}
\includegraphics[width=9cm]{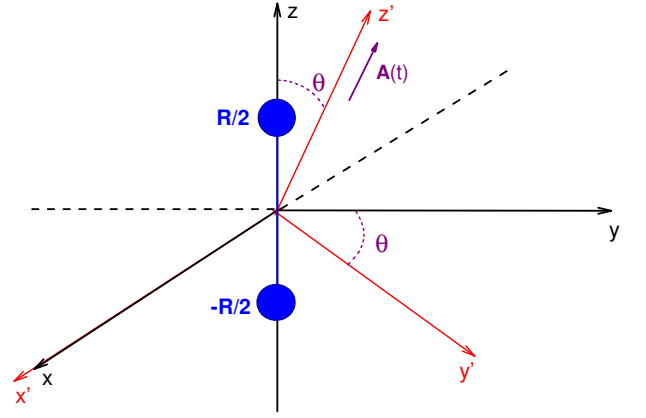}
\end{center}
\caption{Schematic representation of the molecule and laser field frames of
reference, represented by the black and red sets of axis $x,y,z$ and $%
x^{\prime},y^{\prime},z^{\prime}$ respectively. The two centers of the
molecule are apart by $R$ along the $z$ axis of the molecule, and their
positions are indicated by the blue circles in the figure. The field $%
\mathbf{A}(t)$ is polarized along the $z^{\prime}$ axis, and $\protect\theta$
shows the alignment angle of the molecule with respect to the laser field.}
\label{Moleculefigure}
\end{figure}

\begin{eqnarray}
\mathbf{p}_{n} &=&(p_{n||}\cos \theta +p_{n\perp }\sin \theta \sin \varphi )%
\hat{e}_{z}+p_{n\perp }\cos \varphi \hat{e}_{x}  \notag \\
&&\text{ \ }+(p_{n\perp }\cos \theta \sin \varphi -p_{n||}\sin \theta )\hat{e%
}_{y}.  \label{newpcoords}
\end{eqnarray}%
This implies that the momentum components $p_{nx},p_{ny}$ and $p_{nz}$ are
defined by Eq. (\ref{newpcoords}) and that
\begin{equation}
\mathbf{p}_{n}\cdot \mathbf{\mathbf{R}}/2=(p_{n||}\cos \theta +p_{n\perp
}\sin \theta \sin \varphi )R/2.
\end{equation}
A schematic representation of both the field and molecular sets of
coordinates is presented in Fig.~\ref{Moleculefigure}. Below, we
provide the explicit expressions for the integrals
$\mathcal{I}_{n}(n=1,2)$ in the prefactors (\ref{vp1ke1}) and
(\ref{vp2e1}), for the specific types of orbitals employed in this
work.

\subsubsection{Excitation $\protect\sigma \rightarrow \protect\sigma $}

If the second electron is excited from a $\sigma $ to a $\sigma$
orbital, both integrals will have the forms
\begin{eqnarray}
\mathcal{I}_{1} &=&\sum_{j,j^{\prime }}\frac{b_{j}^{(1)}b_{j^{\prime
}}^{(1)}\pi ^{3/2}(-i)^{l_{\alpha }+l_{\beta }}}{2^{^{l_{\alpha }+l_{\beta
}}}(\zeta _{j}+\zeta _{j^{\prime }})^{3/2+l_{\alpha }+l_{\beta }}}  \notag \\
&&\text{ \ \ \ }\times \exp [-\frac{(\mathbf{p}_{1}-\mathbf{k})^{2}}{4(\zeta
_{j}+\zeta _{j^{\prime }})}].\Upsilon (l_{\alpha },l_{\beta })
\end{eqnarray}
\bigskip where%
\begin{equation}
\Upsilon (l_{\alpha },l_{\beta })=\left\{
\begin{array}{c}
1,\text{ \ \ \ }l_{\alpha }+l_{\beta }=0 \\
(\mathbf{p}_{1}-\mathbf{k})_{z},\text{\ \ \ }l_{\alpha }+l_{\beta }=1 \\
2(\zeta _{j}+\zeta _{j^{\prime
}})-(\mathbf{p}_{1}-\mathbf{k})^2_{z},\text{ \ \ }l_{\alpha
}+l_{\beta }=2%
\end{array}%
\right. ,
\end{equation}
and
\begin{equation}
\mathcal{I}_{2}=\sum_{j^{\prime }}b_{j^{\prime }}^{(2)}(-i)^{l_{\beta
}}G(l_{\beta }),  \label{i2sigm}
\end{equation}%
where
\begin{equation}
G(l_{\beta })=\left\{
\begin{array}{c}
2\sqrt{\pi }I_{r}^{(l_{\alpha }=0)},l_{\beta }=0 \\
\left( \tilde{p}_{2z}/\tilde{p}_{2}\right) I_{r}^{(l_{\alpha }=1)},l_{\beta
}=1%
\end{array}%
\right. .  \label{i2sigmsigm}
\end{equation}

In Eq. (\ref{i2sigmsigm}), $I_{r}^{(l_{\alpha }=0)}$ and $I_{r}^{(l_{\alpha
}=1)}$ indicate the radial integrals

\begin{equation}
I_{r}^{(l_{\alpha })}=\int_{0}^{\infty }r^{l_{\beta }+1}j_{l_{\beta }}(%
\tilde{p}_{2}r)\exp [-\zeta _{j}r^{2}]dr,
\end{equation}
where $j_{l_{\beta }}(\tilde{p}_{2}r)$ denotes spherical Bessel
functions.
\subsubsection{Excitation $\protect\sigma \rightarrow \protect\pi $}

We also consider that the second electron is excited from a $\sigma$
orbital to a $\pi$ orbital. In this case, these orbitals are
degenerate. For that reason, we choose to consider a coherent
superposition of the $\pi _{x}$ and $\pi _{y}$ orbitals carrying
equal weights. This gives

\bigskip
\begin{eqnarray}
\mathcal{I}_{1}\hspace*{-0.2cm} &=&\hspace*{-0.2cm}\sum_{j,j^{\prime
}}b_{j}^{(1)}b_{j^{\prime }}^{(1)}\pi ^{3/2}\left[ (-i(\mathbf{p}_{1}-\mathbf{k})_{y})^{l_{%
\beta }}+(-i(\mathbf{p}_{1}-\mathbf{k})_{x})^{l_{\beta }}\right]  \notag \\
&&\text{ \ }\frac{(-i(\mathbf{p}_{1}-\mathbf{k})_{z})^{l_{\alpha
}}}{2^{^{l_{_{\alpha
}}+l_{\beta }}}(\zeta _{j}+\zeta _{j^{\prime }})^{3/2+l_{\alpha }+l_{\beta }}%
}\exp [-\frac{(\mathbf{p}_{1}-\mathbf{k})^{2}}{4(\zeta _{j}+\zeta
_{j^{\prime }})}].
\end{eqnarray}

\bigskip One should note that, if the electron is excited from a $\pi $ to a
$\sigma $ orbital, $\mathcal{I}_{1}$ will also have this form. In the second
prefactor,%
\begin{equation}
\mathcal{I}_{2}=\sum_{j^{\prime }}b_{j^{\prime }}^{(2)}(-i)^{l_{\beta }}%
\left[ \frac{(\tilde{p}_{2y})^{l_{\beta }}+(\tilde{p}_{2x})^{l_{\beta }}}{%
\tilde{p}_{2}}\right] I_{r}^{(l_{\beta })},
\end{equation}%
with $l_{\beta }=1$. Throughout, $(\mathbf{p}_{1}-\mathbf{k})_{\varkappa }$ and $\tilde{p}%
_{2\varkappa },$ with $\varkappa =x,y,z$ are defined according to Eq. (\ref%
{newpcoords}).

\subsection{Interference Condition}

\label{InterferenceCondition}

Here we provide a general interference condition, which takes into account
the structure of the orbitals. This includes $s$ $p$ mixing and the orbital
parity. The expressions that follow are easily derived if the exponentials
in Eqs. (\ref{vp1ke1}) and (\ref{vp2e1}) are expanded in terms of
trigonometric functions. In this case, the prefactor (\ref{vp1ke1}) can be
written as

\begin{equation}
V_{\mathbf{p}_{1}e\mathbf{,k}g}=\frac{V_{12}(\mathbf{p}_{1}-\mathbf{k})}{%
(2\pi )^{3/2}}\sum_{\alpha }\sum_{\beta }\sqrt{C_{+}^{2}-C_{-}^{2}}\sin [\xi
_{1}+(\mathbf{p}_{1}-\mathbf{k})\cdot \mathbf{R}/2]\mathcal{,}
\end{equation}%
with
\begin{equation}
\xi _{1}\mathbf{=}\arctan [\frac{-iC_{+}}{C_{-}}]\text{\ \ \ }
\end{equation}

and%
\begin{equation}
C_{\pm }=1\pm (-1)^{l_{\alpha }+l_{\beta }+\lambda _{\alpha }+\lambda
_{\beta }}.\text{\ }
\end{equation}%
A similar procedure for high-order harmonic generation has been adopted in \cite%
{Dejan2009}. Interference minima are present if

\begin{equation}
\xi _{1}+(\mathbf{p}_{1}-\mathbf{k})\cdot \mathbf{R}/2\mathbf{=}m\pi \text{%
,\ \ }  \label{intrVp1}
\end{equation}

where $m$ is an integer. Similarly, interference maxima are obtained for
\begin{equation}
\xi _{1}+(\mathbf{p}_{1}-\mathbf{k})\cdot \mathbf{R}/2\mathbf{=}(2m+1)\pi /2%
\text{.\ }
\end{equation}%
We will focus on the minima given by Eq. (\ref{intrVp1}) as they are much
easier to observe. If this equation is written in terms of the electron
momentum component $(\mathbf{p}_{1}-\mathbf{k})_{z}$ parallel to the
molecular axis we find%
\begin{equation}
\left[ (p_{1||}-k)\cos \theta +p_{1\perp }\sin \theta \sin \varphi \right]
R/2=m\pi -\xi _{1}.
\end{equation}%
The above-stated equation shows that the parallel momentum component $%
p_{1||} $ parallel to the laser-field polarization will lead to well-defined
interference fringes approximately at
\begin{equation}
p_{1||}=\frac{2(m\pi -\xi _{1})}{R\cos \theta }+k.  \label{intVp1par}
\end{equation}%
This means that, in the plane $p_{1||}p_{2||}$, these minima will be at $%
p_{1||}=const.,$ i.e., parallel to the $p_{1||}$ axis. The perpendicular
component $p_{1\perp }$ will mainly cause a blurring in such fringes, when
the azimuthal angle is integrated over. Extreme limits will be found for
alignment angle $\theta =0$, with sharp two-center patterns, and $\theta
=90^{\circ },$ when they get washed out.

\ Following the same line of argument,%
\begin{equation}
V_{\mathbf{p}_{2}e}=\frac{4\pi }{(2\pi )^{3/2}}\sum_{\alpha }\sqrt{%
D_{+}^{2}-D_{-}^{2}}\sin [\xi _{2}+\mathbf{\tilde{p}}_{2}\cdot \mathbf{R}/2]%
\mathcal{I}_{2},\label{vp2trig}
\end{equation}%
with
\begin{equation}
\xi _{2}\mathbf{=}\arctan [\frac{-iD_{+}}{D_{-}}]\text{\ \ \ }
\end{equation}

and%
\begin{equation}
D_{\pm }=1\pm (-1)^{l_{\beta }+\lambda _{\beta }}\ \text{\ \ .}
\end{equation}
Interference minima are present for Eq. (\ref{vp2trig}) if
\begin{equation}
\xi _{2}+\mathbf{\tilde{p}}_{2}\cdot \mathbf{R}/2\mathbf{=}m\pi
\label{intrVp2}
\end{equation}%
\ \ \ Likewise, there will be interference fringes for
\begin{equation}
\tilde{p}_{2||}=\frac{2(m\pi -\xi _{2})}{R\cos \theta },
\end{equation}%
i.e., parallel to the $p_{2||}$ axis in the plane spanned by the parallel
momentum components $p_{1||},$ $p_{2||}$. In the velocity and the length
gauges, $\tilde{p}_{2||}=p_{2||}$ and $p_{2||}+A(t),$ respectively. Since,
however, $A(t)\simeq 0$ for the electron tunneling time, in practice there
will be very little difference. The perpendicular momentum components will
lead to a blurring in the fringes.

\section{Electron momentum distributions}

\label{results}

In this section, we compute electron momentum distributions, as functions of
the momentum components $(p_{1\parallel },p_{2\parallel })$ parallel to the
laser-field polarization. We assume the external laser field to be a
monochromatic wave linearly polarized along the axis $z^{\prime }$.
Explicitly,
\begin{equation}
\mathbf{E}(t)=\varepsilon _{0}\sin \omega t\hat{e}_{z^{\prime }}.
\end{equation}%
This approximation is reasonable for laser pulses of the order of
ten cycles or longer \cite{X.Liu}. These distributions, when
integrated over the transverse momentum components, read
\begin{eqnarray}
F(p_{1\parallel },p_{2\parallel }) &=&\hspace*{-0.1cm}\iint \hspace*{-0.1cm}%
d^{2}p_{1\perp }d^{2}p_{2\perp }|M_{R}(\mathbf{p}_{1},\mathbf{p}_{2})
\label{distributions} \\
&+&M_{L}(\mathbf{p}_{1},\mathbf{p}_{2})+\mathbf{p}_{1}\leftrightarrow
\mathbf{p}_{2}|^{2},  \notag
\end{eqnarray}%
where $M_{R}(\mathbf{p}_{1},\mathbf{p}_{2})$ and $M_{L}(\mathbf{p}_{1},%
\mathbf{p}_{2})$ refer to the right and left peak in the electron momentum
distributions, respectively, the transition amplitude $M_{R}(\mathbf{p}_{1},%
\mathbf{p}_{2})$ is given by Eq.~(\ref{Mp}), and $d^{2}p_{n\perp }=$ $%
p_{n\perp }dp_{n\perp }d\varphi _{p_{n}}.$ For a monochromatic field, we can
use the symmetry $\mathbf{A}(t)=-\mathbf{A}(t\pm T/2),$ where $T=2\pi
/\omega $ corresponds to a field cycle, in order to simplify the computation
of \ the electron momentum distributions. This is explained in detail in our
previous work \cite{RESIM}. We also symmetrize the above-stated
distributions with respect to the particle exchange $\mathbf{p}%
_{1}\leftrightarrow \mathbf{p}_{2}.$ To a good approximation, it is
sufficient to consider the incoherent sum in Eq. (\ref{distributions}) as
the interference terms between the right and left peaks practically get washed out upon
the transverse momentum integration (see Appendix B in \cite{RESIM}).

In the following, we will compute electron momentum distributions for $%
\mathrm{Li}_{2}$ and $\mathrm{N}_{2}.$ For all cases, we assume that
the
electron-electron interaction is of contact type, i.e., $V_{12}=\delta (%
\mathbf{r}_{1}-\mathbf{r}_{2}).$ This will avoid a further momentum bias in
the electron-electron distributions as it leads to $V_{12}(\mathbf{p}_{1}-%
\mathbf{k})=const.$ and allow us to investigate the influence of the target
structure alone. For a long-range potential, $V_{12}(\mathbf{p}_{1}-\mathbf{k%
})$ would be momentum dependent, and hence mask the features we
intend to investigate.

\subsection{Interference effects and s p mixing}

\label{alignment}

We will commence by investigating whether the interference
conditions derived in Sec. \ref{InterferenceCondition} hold. For
that purpose, we must have non-negligible tunneling ionization for
parallel-aligned molecules, as this is the situation for which the
fringes are expected to be sharpest. Hence, one must consider a
target for which neither the HOMO nor the LUMO exhibits nodal planes
along the internuclear axis. Therefore, we assume that the first
electron tunnels from the HOMO in $\mathrm{Li}_2$ and rescatters
inelastically with $\mathrm{Li}_{2}^{+},$
exciting the second electron from its HOMO (2$\sigma _{g})$ to its LUMO\ (2$%
\sigma _{u})$. In order to get a clear picture of conditions
(\ref{intrVp1}) and (\ref{intrVp2}), we must investigate the
corresponding prefactors individually.

\begin{figure}[th]
\begin{center}
\hspace*{-1cm}\includegraphics[width=11.5cm]{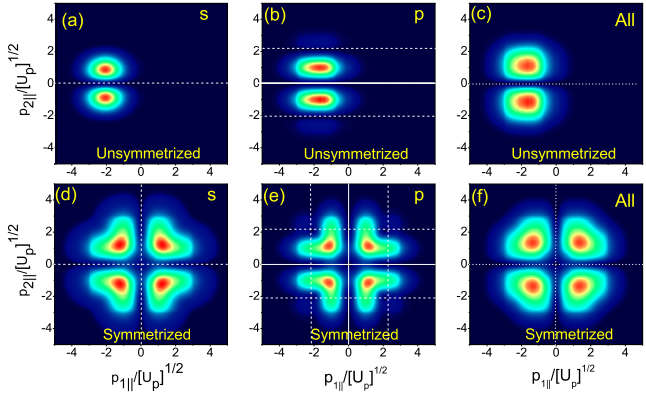}
\end{center}
\caption{Electron-momentum distributions for NSDI in
$\mathrm{Li}_{2}$ (bound-state energies $E_{1g}$ $=$
$0.18092040$ a.u., $E_{2g}$ $=$ $0.43944428$ a.u. and $E_{2e}$ $=$ $%
0.12481836 $ a.u. and equilibrium internuclear distance $R=4.7697$
a.u.) considering only the RESI mechanism, as functions of the
momentum components parallel to the laser-field polarization$,$
obtained considering $V_{\mathbf{p}_{2}e}$
according to Eq. (\protect\ref{Vp2e}) and $V_{\mathbf{p}_{1}e,\mathbf{k}%
g}=const$. We consider zero alignment angle, driving-field intensity $%
I=4.6\times 10^{13}\mathrm{W/cm}^{2}$ and $\protect\omega =0.057$
a.u. respectively. Panels (a) to (c) display only the contribution
from the orbits starting in the first half cycle of the field, while
in panels (d) to (f) the distributions have been symmetrized to
account for the electron orbits starting in the other half cycle and
for electron indistinguishability. The left, middle and right panels
correspond to the contributions of the $s$, $p$ and all states used
in the construction of the $\protect\sigma _{u}$ LUMO, respectively.
The solid, dashed and short dashed lines show the position of minima
due to the two-center interference, node of the wavefunction and
mixed cases, respectively. The contour plots have been normalized to
the maximum probability in each panel. } \label{LUMO0deg}
\end{figure}

In Fig.~\ref{LUMO0deg}, we depict the above-mentioned electron-momentum
distributions for alignment angle $\theta =0^{\circ}.$ We consider $V_{\mathbf{p}%
_{1}e\mathbf{,k}g}=const.$ and focus on the influence of $V_{\mathbf{p}%
_{2}e} $ alone. We take either the individual contributions of $s$
and $p$ states or the combination of both\ for 2$\sigma _{u}$. For
clarity, in the upper panels, we also exhibit the distributions
obtained without symmetrizing with respect to the momentum exchange
and electron start times. For all cases, the two-center fringes in
Fig.~\ref{LUMO0deg} are parallel to $p_{2||}=const.$, in agreement
with the second interference condition derived in
Sec.~\ref{InterferenceCondition}.

For pure $s$ or $p$ states and $\lambda _{\alpha }=1$, which is the
case for a $\sigma_u $ orbital, this condition can be further
simplified. It
reduces to%
\begin{equation}
\sin [\mathbf{\tilde{p}}_{2}\cdot \mathbf{R}/2]=0,
\end{equation}%
for $s$ states, and

\begin{equation}
\cos [\mathbf{\tilde{p}}_{2}\cdot \mathbf{R}/2]=0
\end{equation}
for $p$ states. This implies that, for the former, we expect minima at $%
\mathbf{\tilde{p}}_{2}\cdot \mathbf{R}=2m\pi ,$ while for the latter they
should occur at $\mathbf{\tilde{p}}_{2}\cdot \mathbf{R}=(2m+1)\pi .$ The
position of such minima can also be determined analytically by considering
that the second electron tunnels at the peak of the laser field, i.e., at $%
t=\pi /2$. The dashed lines in the figure show that the position of these
minima exhibit a very good agreement with this simple estimate. Physically,
this good agreement may be attributed to the fact that the second electron
tunnels most probably at this time.

For the $s$ states the two-center interference gives a sharp minimum at $%
p_{2\parallel }=0$ (Figs.~\ref{LUMO0deg} (a) and (d)), while for the
$p$ states these patterns are located near $p_{2\parallel }=\pm
3\sqrt{U_{p}}$ (Figs.~\ref{LUMO0deg} (b) and (e)). In the $p-$state
case the distribution has another minimum at $p_{2\parallel }=0,$
which comes from the fact the $p$ wavefunctions vanish for
$\mathbf{p}_n=0$. This causes a suppression in the transition
amplitude. If the contributions of both $s$ and $p$ states are
considered, the minima in the high-momentum region due to the
two-center
interference seen for the $p$ states vanish, but the minimum at $%
p_{2\parallel }=0$ survives. This is shown in Figs.~\ref{LUMO0deg}.(c) and
(f) for unsymmetrized and symmetrized distributions, respectively.

One should note, however, that for parallel-aligned molecules, both
the two-center minimum for the $s$ states and the minimum caused by
the node in the $p$ states occur at the same momentum, i.e., at
$p_{2\parallel }=0$. Hence, when $s$ $p$ mixing is included both
mechanisms contribute to the suppression at the axes
$p_{n\parallel}=0$ seen in Figs.~\ref{LUMO0deg}.(c) and (f). We will
now investigate the behavior of this node when the alignment angle
is varied. Since for Li$_{2}$ both the LUMO and the HOMO exhibit
distinct shapes and symmetries one can expect significant changes in
the electron-momentum distributions when this angle is modified.

\begin{figure}[t]
\begin{center}
\includegraphics[width=8.5cm]{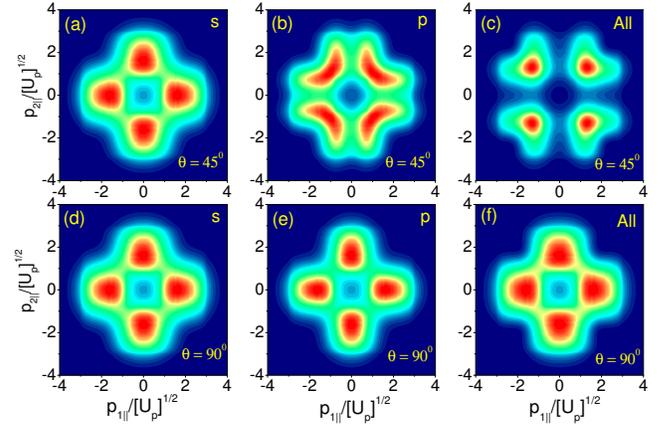}
\end{center}
\caption{Electron-momentum distributions for RESI in $\mathrm{Li}_{2}$ as functions
of the electron momentum components parallel to the laser-field polarization
considering $V_{\mathbf{p}_{1}e,\mathbf{k}g}=const$ and $V_{\mathbf{p}_{2}e}$
according to Eq. (\protect\ref{Vp2e}), for alignment angles $\protect\theta %
=45^{0}$ (panels (a) to (c)), and $90^{0}$ (panels (d) to (f)). The
remaining parameters are the same as in the previous figures. The solid
lines show the position of minima due to the node of the one-center
wavefunction. From left to right, we considered the contributions of the $s$%
, $p$ and all states used in the construction of the LUMO. All panels have
been symmetrized with regard to the electron orbits and
indistinguishability. The contour plots have been normalized to the maximum
probability in each panel. }
\label{LUMO4590deg}
\end{figure}

Hence, in Fig.~\ref{LUMO4590deg}, we consider the same prefactors as
in the previous case, but alignment angles $\theta =45^{0}$ and
$90^{0}$. The figure shows that the patterns caused by the electron
emission at spatially separated centers get washed out for such
angles. This is due to the momentum components perpendicular to the
laser-field polarization, and can be seen very clearly in
Fig.~\ref{LUMO4590deg}.(a), where the $s$ contributions are
displayed for $\theta=45^{\circ}$. Already for this angle the
interference minima at the axes $p_{n\parallel}=0$ are absent. In
contrast, the suppression at the axes caused by the fact that the
$p$ wavefunctions vanish in that momentum region is still present.
This is shown in Fig.~\ref{LUMO4590deg}.(b), in which the
contributions from the $p$ states are depicted. The blurring is
caused by the fact that, in momentum space, these wavefunctions are
proportional to $G(l_{\beta}=1)$ (see Eq.~(\ref{i2sigmsigm})). This
function contains components of $\mathbf{p}_2$ both parallel and
perpendicular to the laser field polarization, and the contributions
from the latter tend to wash out the minimum. When both $s$ and $p$
contributions are considered, there is a strong suppression of the
yield near the $p_{n\parallel}=0$ axis (see
Fig.~\ref{LUMO4590deg}.(c)). We have verified that this is due to
the destructive interference between both types of contributions in
this momentum region.

For $\theta=90^{\circ}$, only the components $p_{2\perp}$
contribute, and the electron momentum distributions are determined
by the momentum-space integration alone. As a result, they reflect
the momentum-space constraints for the RESI mechanism. These
constraints lead to electron momentum distributions peaked at
$(p_{i\parallel},p_{j\parallel})=(\pm 2\sqrt{U_p},0)$, with
$i,j=1,2$ and $i\neq j$ and with widths $2\sqrt{U_p}$, and have been
explicitly written in \cite{Shaaran,RESIM}. This holds both for the
$s$, $p$ and mixed case (Figs.~\ref{LUMO4590deg}.(d), (e) and (f),
respectively).

\begin{figure}[t]
\begin{center}
\hspace*{-1cm}\includegraphics[width=10cm]{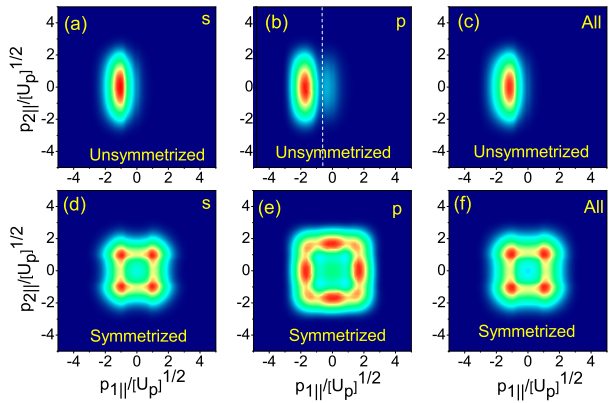}
\end{center}
\caption{RESI electron-momentum distributions for $\mathrm{Li}_2$ considering $V_{\mathbf{p}%
_{2}e}=const.$ and $V_{\mathbf{p}_{1}e,\mathbf{k}g}$ according to Eq. (%
\protect\ref{vp1ke1}), for $\protect\theta =0$. The field and molecular
parameters are the same as in the previous figure. The upper panels display
only the contribution from the sets of orbits starting in the first half
cycle of the laser field. In the lower panels the distributions have been
symmetrized in order to account for the orbits starting in the other half
cycle of the field, and for electron indistinguishability. The left, middle
and right panels display the contributions from $s$, $p$ and all states
composing the HOMO and the LUMO, respectively. The dashed line shows the
position of the two-center interference minimum. The contour plots have been
normalized to the maximum probability in each panel. }
\label{HOMO0deg}
\end{figure}

We will now focus on the interference condition determined by the excitation
prefactor (\ref{Vp1e,kg}). With this objective, we will keep $V_{\mathbf{p}%
_{2}e}=const.$ and investigate the influence of $V_{\mathbf{p}_{1}e,\mathbf{k%
}g}$ alone, starting from vanishing alignment angle. Once more, we will
study the contributions of the $s$ and $p$ states, and the overall
distributions. The interference condition and also the wavefunctions in the
excitation prefactor now incorporate the HOMO and the LUMO (see Eq.~(\ref{Vp1e,kg})). For $\mathrm{Li}%
_{2}^{+},$ the former and the latter are a gerade and an ungerade orbital,
so that $\lambda _{\alpha }=0$ and\ $\lambda _{\beta }=1$ in Eq.~(\ref{intrVp1}%
). For pure $s$ states, $l_{\alpha }=l_{\beta }=0$ and for pure $p$ states, $%
l_{\alpha }=l_{\beta }=1.$ This will lead to the simplified interference
condition
\begin{equation}
\sin [\mathbf{(p}_{1}-\mathbf{k)}\cdot \mathbf{R}/2]=0
\end{equation}%
for both. Hence, one expects a minimum close to vanishing parallel momenta
in the pure cases. When $s$ $p$ mixing is included, however, different
angular momenta will also be coupled and the general interference condition
must be considered.

The electron momentum distributions obtained in this way are shown
in Fig.~\ref{HOMO0deg}, for both symmetrized and unsymmetrized
distributions (upper and lower panels, respectively). For most
distributions in the figure, we do not observe a clear suppression
of the probability densities in any momentum region. This holds both
for those caused by the two center interference and by the geometry
of the wavefunctions at the ions. We have only observed a two center
minimum if we consider the individual contributions of the $p$
states, and do not symmetrize the distributions (see Fig.
\ref{HOMO0deg}.(b)). This is due to the fact that, for the
parameters considered in this work, the two-center minimum according
to condition (\ref{intVp1par}) lies at or beyond the boundary of the
momentum region for which rescattering of the first electron has a
classical counterpart. The center of this region is roughly at
$p_{1||}\simeq 2\sqrt{U_{p}}$ and its extension is determined by the
difference between the maximal electron kinetic energy upon return
and the excitation energy $E_{2g}-E_{2e},$ as discussed in our
previous article \cite{RESIM}.

Apart from that, $s$ $p$ mixing will lead to a blurring of this
minimum, as it couples states with different angular momenta.
Symmetrization introduces other events, either due to the electron
indistinguishability or displaced by half a cycle, and obfuscates
this minimum further, as shown in the lower panels of the figure.

If the alignment angle is varied, incorporating only the excitation prefactor $V_{\mathbf{p}_1e,\mathbf{k}g}$ will lead to
ring-shaped distributions, 
regardless of whether only $p$, $s$ or all basis states employed in
the construction of the HOMO and LUMO are taken. This is expected
as, apart from the above-mentioned $s$ $p$ mixing, which blur
wavefunction-specific features, there will now be transverse
momentum components in the prefactor $V_{\mathbf{p}_1e,\mathbf{k}g}$
which will wash out two-center interference patterns. We have
verified that this is indeed the case (not shown).

\subsection{Molecular orbital signature}

\label{orbits}

In this section, we make an assessment of how the geometry of the
HOMO and the LUMO affect the RESI electron momentum distributions.
With this objective, we incorporate both prefactors
$V_{\mathbf{p}_{2}e}$ and $V_{\mathbf{p}_{1}e,\mathbf{k}g}$ and vary
the alignment angle. In order to discuss the influence of nodal
planes, we are also providing the overall yield obtained in our
computation in the two figures that follow (see color maps on the
right-hand side of each panel). From other strong-field phenomena,
it is well-known that the presence of nodal planes may suppress the
yield considerably \cite{Dejan2009,CarlaBrad}.
\begin{widetext}
\begin{figure}[h]
\begin{center}
\hspace*{-0.5cm}\includegraphics[width=16cm]{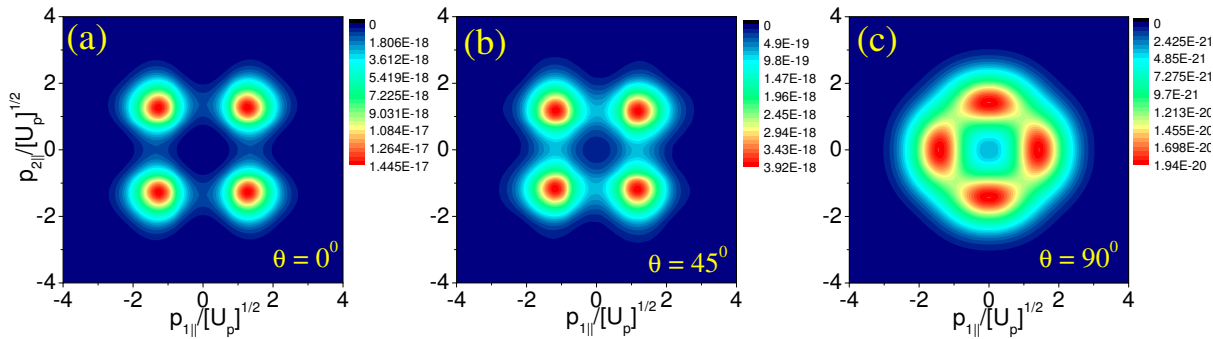}
\end{center}
\vspace*{-0.5cm} \caption{Electron-momentum distributions for Li$_2$
as functions of the parallel momenta $(p_{1\parallel },p_{2\parallel
})$ considering all prefactors, for different alignment angles.
Panel (a), (b) and (c) correspond the alignment angle $\theta$= 0,45
and 90 degrees, respectively. The field and molecular parameters are
the same as in the previous figures.}\label{Li2HOMOLUMOall}
\end{figure}
\end{widetext}

We will commence by having a closer look at $\mathrm{Li}_2$. Such
results are displayed in Fig.~\ref{Li2HOMOLUMOall}. The main
conclusion to be drawn from the figure is that the prefactor
$V_{\mathbf{p}_{2}e}$ plays the dominant role in determining the
shapes of the electron momentum distributions. This can be observed
by a direct comparison of Fig.~\ref{Li2HOMOLUMOall}.(a) with Fig.
\ref{LUMO0deg}.(f), for vanishing alignment angle. The distributions
in both figures exhibit similar shapes and minima at the axes
$p_{n||}=0$, and are very different from those obtained if only the
recollision-excitation prefactor
is included (see Fig.~\ref{HOMO0deg}(f)). The main effect of the excitation prefactor $V_{\mathbf{p}%
_{1}e,\mathbf{k}g}$ is to alter the widths of the distributions. This
situation persists for larger angles, such as $\theta =45^{\circ }$ and $%
\theta =90^{\circ },$ as a comparison of
Figs.~\ref{Li2HOMOLUMOall}.(b) and (c), with
Fig.~\ref{LUMO4590deg}.(c) and (f) shows. For $\theta=45^{\circ}$,
there is a suppression of the yield near the axes
$p_{n\parallel}=0$, while for $\theta=90^{\circ}$ the interference
patterns are washed out.

Another interesting feature is that the overall yield decreases with
the alignment angle between the molecular axis and the field. This
is due to the fact that the LUMO, from which the second electron
tunnels, is a $\sigma$ orbital. Spatially, $\sigma$ orbitals are
localized along the internuclear axis, and do not exhibit nodal
planes for vanishing alignment angle. This implies that tunneling
ionization is favored when the LUMO is parallel to the laser field,
and decreases when the difference between the orientation between
the field and the LUMO increases.

A legitimate question is, however, how the shape of the molecular
orbital to which the second electron is excited is imprinted on the
electron momentum distribution, if there are nodal planes parallel
or perpendicular to the molecular axis. For that reason, we now
present electron momentum distributions under the assumption that
the second electron is excited to a $\pi_g$ orbital. Specifically,
we choose $\mathrm{N}_2$ and its singly ionized counterpart, i.e.,
$\mathrm{N}^+_2$ as the molecular species in our RESI computation.
The first electron will be ripped off from the HOMO, which is a
$3\sigma_g$ orbital. However, upon return, it will excite the second
electron to the LUMO, which is a $1\pi_g$ orbital. A $1\pi_g$
orbital exhibits two nodal planes, which will be oriented along the
laser-field polarization for parallel and perpendicular-aligned
molecules, i.e., at alignment angles $\theta=0^{\circ}$ and
$\theta=90^{\circ}$. This orbital also exhibits lobes at angles
$\theta=(2n+1)\pi/4$ with regard to the internuclear axis.

\begin{widetext}
\begin{figure}[t]
\hspace*{-1.5cm}\includegraphics[width=16cm]{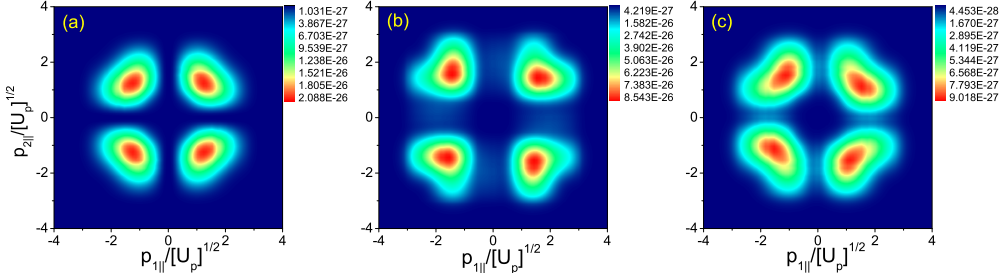}
\caption{Electron-momentum distributions for $\mathrm{N}_2$
(bound-state energies $E_{1g}=0.63486$ a.u., $E_{2g}=1.12657$ a.u.,
and $E_{2e}=0.26871290$ a.u. and equilibrium internuclear distance
$R=2.11$ a.u.) in a linearly polarized monochromatic field of
intensity $I=1.25 \times 10^{14}\mathrm{W}/\mathrm{cm}^2$ as
functions of the parallel momenta $(p_{1\parallel },p_{2\parallel
})$ considering all prefactors, for alignment angles $\theta=0$,
$45$ and $90$ degrees(panels (a), (b) and (c), respectively).}
\label{N2HOMOLUMO}
\end{figure}
\end{widetext}

The results obtained for this molecular species are exhibited in
Fig.~\ref{N2HOMOLUMO}. As an overall pattern, we observe that the
NSDI signal no longer decreases monotonically with increasing
alignment angle. In fact, the signal increases for alignment angle
$0<\theta<45^{\circ}$, is strongest for $\theta=45^{\circ}$, and
decreases once more for $45^{\circ}<\theta<90^{\circ}$. This may be
easily understood as a consequence of the geometry of the $1\pi_g$
orbital. For $\theta=0^{\circ}$ (Fig.~\ref{N2HOMOLUMO}(a)), the
external field is parallel to one of the nodal planes. Hence, tunnel
ionization is strongly suppressed. This reflects itself in the
overall yield. As the alignment angle increases, the
field-polarization direction gets further and further away from the
direction of this nodal plane, and the yield increases until
$\theta=45^{\circ}$ (Fig.~\ref{N2HOMOLUMO}(b)). For this angle, the
field is parallel to one of the lobes of the $1\pi_g$ orbital, so
that tunnel ionization of the second electron is enhanced. As the
alignment angle is further increased, the direction of the field
approaches the nodal plane at $\theta=\pi/2$ and ionization is
further suppressed (Fig.~\ref{N2HOMOLUMO}(c)).

Apart from the above-mentioned  behavior, we also observe a
suppression along the axes $p_{n\parallel}=0$, regardless of the
alignment angle. This is due to the fact that, in position space,
$\pi$ orbitals vanish at the origin of the coordinate system.
Consequently, their Fourier transform vanish for $p_{n\parallel}=0$.
\section{Conclusions}
\label{conclusions}

The results presented in the previous sections illustrate the
potential of laser-induced nonsequential double ionization for the
attosecond imaging of molecules. This is particularly true if the
recollision-excitation with subsequent tunneling ionization (RESI)
pathway is dominant. The computations in this work show that the
shapes of the RESI electron momentum distributions depend in a
dramatic fashion on the geometry of the state to which the second
electron has been excited by the first electron, and from which it
tunnels. The state in which the second electron is initially bound,
i.e., the highest occupied molecular orbital (HOMO) of the singly
ionized species plays only a secondary role.

Thereby, two main issues are important in determining the shapes of
the electron momentum distribution: the quantum interference caused
by the interference due to the photoelectron emission at spatially
separated center, and the geometry of the orbital from which the
second electron tunnels.

In order to investigate the first issue, generalized interference
conditions for the first and second electron that take into account
$s$ $p$ mixing along the lines of \cite{Dejan2009} have been
derived, and led to fringes parallel to the $p_{n\parallel}=0$
$n=1,2$ axes in the plane spanned by the electron momentum
components parallel to the laser-field polarization. These fringes
agreed well with analytic estimates, but were washed out for
relatively small alignment angles.

In contrast, the features caused by the orbital geometry, such as
suppression of the probability density near $p_{n\parallel}=0$
observed for $p$ states were present over a wide range of alignment
angles. Furthermore, the presence or absence of nodal planes
manifests itself as the suppression, or enhancement, of the overall
yield with regard to the alignment angle.  We have discussed the
differences and similarities between $\sigma_u$ and $\pi_g$ orbitals
in this context, exemplified by the LUMOs of  $\mathrm{N}_2$ and
$\mathrm{Li}_2$. These results agree with those reported in the
literature for phenomena such as high-order harmonic generation
\cite{CarlaBrad,Olga2009,Haessler2010} and above-threshold
ionization \cite{Lin}.

\noindent \textbf{Acknowledgements:} This work has been financed by
the UK EPSRC (Grant no. EP/D07309X/1) and by the STFC. We are
grateful to P. Sherwood, J. Tennyson and M. Ivanov for useful
comments, and to M. T. Nygren for his collaboration in the early
stages of this project. C.F.M.F. and B.B.A. would like to thank the
Daresbury Laboratory for its kind hospitality.

\end{document}